\documentclass[11pt]{article}
\usepackage[latin1]{inputenc}
\usepackage[english]{babel}
\usepackage[namelimits]{amsmath}
\usepackage{authblk}
\usepackage{amssymb}
\usepackage{amsmath}
\usepackage{amsthm}

{\tiny }

\begin{document}
\title{{\bf The cosmological constant from Planckian fluctuations and the averaging procedure}}
\author[1,2,3]{S. Viaggiu\thanks{viaggiu@axp.mat.uniroma2.it and s.viaggiu@unimarconi.it}}
\affil[1]{Dipartimento di Fisica Nucleare, Subnucleare e delle Radiazioni, Universit\'a degli Studi Guglielmo Marconi, Via Plinio 44, I-00193 Rome, Italy.}
\affil[2]{Dipartimento di Matematica, Universit\`a di Roma ``Tor Vergata'', Via della Ricerca Scientifica, 1, I-00133 Roma, Italy.}
\affil[3]{INFN, Sezione di Napoli, Complesso Universitario di Monte S. Angelo,
	Via Cintia Edificio 6, 80126 Napoli, Italy.}

\date{\today}\maketitle
\begin{abstract}
\noindent In this paper I continue the investigation in \cite{1,1b} concerning my proposal on the nature of the cosmological constant.
In particular, I study both mathematically and physically the quantum Planckian context and I provide, in order to depict
quantum fluctuations and in absence of a 
complete quantum gravity theory, a semiclassical solution where an effective inhomogeneous metric at Planckian scales or above
is averaged. 
In such a framework, a generalization of the well known Buchert formalism
\cite{2} is obtained with the foliation in terms of the mean value $s(\hat{t})$ of the time operator $\hat{t}$ in a maximally localizing state $\{s\}$ of a quantum spacetime \cite{3,4,5,6} and in a cosmological context \cite{7}. As a result, after introducing a 
decoherence length scale $L_D$ where quantum fluctuations are averaged on, a classical de Sitter universe emerges 
with a small cosmological constant depending on $L_D$ and frozen in a true vacuum state (lowest energy), provided that the kinematical backreaction is negligible at that scale $L_D$. 
Finally, I analyse the case with a non-vanishing initial spatial curvature
$\mathcal{R}$ showing that, for a reasonable large class of models, spatial curvature and kinematical backreation 
$\mathcal{Q}$ are suppressed by the dynamical evolution of the spacetime.
 
\end{abstract}

\section{Introduction}

Vacuum Planckian fluctuations at the Planck length scale $L_P$
are generally expected to generate a cosmological constant \cite{8,9,10,11,12,13,14}
$\Lambda$ looking like $\Lambda\sim 1/L_P^2$. This is a rather huge value, about $10^{122}$ greater than the one effectively
observed, thus representing an embarassing problem from a theoretical point of view. In order to be in agreement with the value for $\Lambda$ dictated by quantum field theory, an improbable fine tuning of about $122$ orders is required. In such a 
irrealistic case, supersymmetry is required, but this elegant mechanism has not been observed at LHC collider. New ideas are thus
urgent. To this purpose, an alternative approach to the usual one can be found in \cite{15}, where a semiclassical approximation is considered with 
quantum fluctuations generating a stochastic field depicted in terms of an effective inhomogeneous metric. There, a de Sitter universe
with a small cosmological constant emerges, after invoking parametric resonance. Another new study can be found in \cite{16}
with the introduction of a dynamical cosmological constant embedded in a background obtained by an extension of the general relativity
in terms of the Ashtekar variables. The authors of \cite{16} found a new possible uncertainty relation between the dynamical
cosmological constant and the Chern-Simons time. In \cite{17} the author suggests that the cosmological constant is effectively
of the order of $1/L_P^2$, but it is hidden by quantum fluctuations that in turn generate an inhomogeneous spacetime, both in time and space. The author claims that its proposal is a practical realization of the old Wheeler idea \cite{18} concerning the spacetime 
at the Planckian scales seen as a 'spacetime foam'. The interesting feature of the study presented in \cite{17} is that in practice the strong inhomogeneities of the spacetime at Planckian scales inhibit dynamic in the spacetime, with a large class of initial data 
hiding the effects of a huge $\Lambda$ at a macroscopic level within the Buchert averaging scheme. Many technical and physical issues are present in \cite{17}, as for example the definition of a suitable time coordinate. However, the model in \cite{17} cannot explain the smallness of the cosmological constant. Finally, my proposal in \cite{1} is\footnote{See also \cite{19,20,21} for an application
of my proposal to the black hole case and \cite{22} in a more general context and \cite{22a} for an earlier proposal also in terms
of massless excitations within the apparent horizon.}
an attempt to follow a physically and 
mathematically sound alternative point of view. In \cite{1} the cosmological constant $\overline{\Lambda}$
is splitted in the usual way as $\overline{\Lambda}=\Lambda+{\Lambda}_{vac}$, where $\Lambda$ is the non-interacting bare
cosmological constant and ${\Lambda}_{vac}$ represents the contribution due to quantum fluctuations, with
$\overline{\Lambda}$ the observed one. The new idea in \cite{1} is that Planckian fluctuations can generate a very large 
(interacting) cosmological  constant, due to large density fluctuations, but these fluctuations average on bigger and bigger scales and as a result the effective cosmological constant becomes smaller, up to the decoherence length scale $L_D$ where a de Sitter 
spacetime emerges with $\overline{\Lambda}$ frozen in the lowest energy state. In such a way my model can explain the birth
of the cosmological constant in terms of averaged Planckian fluctuations and its smallness by introducing a quantum decoherence scale.
Within a semiclassical model, a modified Buchert averaging scheme is used. In this paper I further analyse the mathematical and physical background in \cite{1}, paying particular attention to the averaging procedure.\\ 
In section 2 we specify the classical background together with the prescription to obtain, generalizing the theorem in \cite{1},
the cosmological constant equation of state. In section 3 Planckian fluctuations
are studied  by means of a quantum spacetime,
while in section 4 a semiclassical solution with an averaging procedure is presented. Finally , section 5 collects conclusions and 
final remarks. 
  
\section{Classical background and the equation of state for $\Lambda$}

In this paper I depict the genesis of a positive cosmological constant. Hence, without loss of generality, as also stated in
\cite{1}, it is sufficient to consider as a classical background the de Sitter spacetime with zero spatial
curvature in comoving coordinates:
\begin{equation}
ds^2=-c^2dt^2+a^2(t)\left(dr^2+r^2d\theta^2+r^2\sin^2\theta\;d\phi^2\right),
\label{2}	
\end{equation}
where $a(t)\sim e^{ct\sqrt{\frac{\overline{\Lambda}}{3}}}$ and $\overline{\Lambda}$ is the measured (interacting)
cosmological constant.
To start with, we consider a spherical ball of proper areal radius $L=a(t)r$.
As shown in \cite{1}, the first step of my approach is to realize that the matter energy content within $L$ is not arbitrary
and quasi-local energy $E_{ms}=M_{ms}c^2$, i.e. the quasi local Misner-Sharp mass $M_{ms}$ \cite{23}, can be calculated 
at a classical level, to obtain 
$E_{ms}=\frac{c^4}{2G}\frac{L^3}{L_A^2}$, where $L_A^2=\frac{3}{\overline{\Lambda}}$. The first
goal is to obtain a physical microscopic description 
of the cosmological constant. The new approach presented in \cite{1} is based on three fundamental
considerations, often missing in the literature.\\
To start with, in the usual treatment of the cosmological constant, the vacuum energy is obtained as a summation over all possible vacuum contributions from different energy scales. However, it should be stressed that 
in the energy-momentum tensor $T_{\mu\nu}$ energies composing the universe enter as densities. A vacuum energy 
representing a cosmological constant with the suitable 
equation of state can be written as $T_{\mu\nu}=-8\pi G{\rho}_{vac} g_{\mu\nu}$. Hence, when one considers a given specific contribution to
the vacuum density ${\rho}_{vac}$, it should be specified, in a general relativistic context, the volume, i.e. the length scale,
at which a given energy density emerges. This is a crucial point in my new approach to the cosmological constant problem.\\
A second important ingredient of the proposal in \cite{1}, as mentioned above, 
concerns the geometrical constraints on the energy provided by 
general relativity. In fact, if one considers, for example, a spherical region of areal radius $L$, general relativity furnishes
the quasi-local energy within $L$ in terms of the Misner-Sharp mass \cite{23}. As a consequence, if we want to depict the cosmological constant in terms of energy densities in a given volume, then this constraint must be taken into account. 
In this form, this constraint is practically absent in the literature. In fact, in the usual approach 
present in quantum field theory, energy densities $\rho$ are expressed in terms of the expectation value of 
$T_{\mu\nu}$ in some state $\{s\}$. Unfortunately, we have not at our disposal a complete universally 
accepted quantum gravity theory, with a complete non-commutative spacetime at Planckian scales for a curved spacetime.
In light  of these deficiencies, we use a semiclassical 
phenomenological approach: the metric is assumed to be classical, but coordinates $\{x^{\alpha}\}$ belong to the spectrum of quantum operators $\{\hat{x^i}\}$ satisfying physically motivated spacetime uncertainty relations
(STUR). To this purpose, in the classical background and for a 
Friedmann spacetime, the Misner-Sharp mass in a proper volume $V$
is provided by ${\rho}_{ms} V$. Hence, for the reasonings above, we must obtain an expression for the energy density from
semiclassical reasonings of the form $\rho={\rho}_{ms}+{\rho}_{f}$, where ${\rho}_f$ is provided by Planckian fluctuations.
These fluctuations are expected to be very strong at Planckian scales and negligible at scales well above the Planck one.\\
Another important point refers to the equation of state of the cosmological constant. As it is
well known, the equation of state for 
the cosmological constant is $c^2{\rho}_{vac}=-{p}_{vac}$, with $p_{vac}$ the pressure and 
$\{{\rho}_{vac}, p_{vac}\}$ constant both in time and space. As an example, 
if one considers the usual expression of the energy levels of an harmonic oscillator, 
$E_n=\left(n+\frac{1}{2}\right)\hbar\omega$, with $\omega$ the angular frequency, the vacuum energy, according
to Heisenberg uncertainty principle, is $E_0\neq 0$. However, this contribution to the vacuum energy is related to a radiation
field, rather than to one with a cosmological-like equation of state. This simple reasoning indicates that a vacuum energy
effectively contributes to the cosmological constant if and only if the equation of state $c^2{\rho}_{vac}=-{p}_{vac}$ is 
satisfied: only in this case we can write $T_{\mu\nu}=-8\pi G{\rho}_{vac} g_{\mu\nu}$.
My new idea presented in \cite{1} is that the vacuum contributions considered in the literature
(QED, QCD, GUT energies...) in practice represent radiation fields and do not contribute to an effective cosmological 
constant equation of state or more generally to an inflationary universe. 
Only Planckian fluctuations can generate a cosmological constant equation of state. The physical mechanism,
depicted also in \cite{21} for a black hole and generalized in \cite{22}, is capable to transform a radiation field into one with
a $\gamma$ linear equation of state with ${p}_{vac}=\gamma c^2{\rho}_{vac}$. 
This can be done by considering the cosmological constant composed of massless excitations where Planckian fluctuations come into action to transform the initial radiation-like equation of state (with $\gamma=1/3$) into a cosmological constant one ($\gamma=-1$). 
To this regard, we can generalize the proposition shown in \cite{1}.
To be more precise, suppose to have a radiation field composed of massless excitations with frequency spectrum 
$\{\omega^{(0)}\}$, with partition function $Z^{(0)}$, internal energy $U^{(0)}>0$, free energy
$F^{(0)}$ and equation of state 
$P^{(0)}=c^2\frac{\rho^{(0)}}{3}$. Note that the spectrum $\omega^{(0)}$ can be discrete as in \cite{1} or also 
continuous: in any case the following proposition still holds:\\
\noindent {\bf Proposition:} {\it Let ${\omega}^{(0)}$ denote the angular frequency of $N$ massless excitations within a volume $V$ of proper areal radius $L$. The excitations with energy
$\hbar\omega=\hbar{\omega}^{(0)}+\hbar\frac{\Phi(L)}{N}$ have a linear equation of state
$PV=\gamma U$ provided that the differentiable function $\Phi(L)$
satisfies the following equation} 
\begin{equation}
\hbar\left[L\;{\Phi}_{,L}(L)+\Phi(L)\right]=U(L)(1-3\gamma), \label{3}
\end{equation}
{\it together with the condition}
\begin{equation}
U(L)-\hbar\;\Phi(L) > 0.
\label{4}
\end{equation}
\begin{proof}
With the usual relation $U(L)=-{\ln(Z_T)}_{,\beta}$, we obtain
\begin{equation}
U=U^{(0)}+\hbar\;\Phi(L).
\label{5}
\end{equation}	
Since we have $U^{(0)}>0$, condition (\ref{5}) follows. For the free energy $F$ we have
$F=-N K_B T\ln(Z)=F^{(0)}+\hbar\;\Phi(L)$. Moreover
\begin{equation}
{F}_{,V}=\hbar\;{\Phi}_{,L}\;L_{,V}+L_{,V}\;{F^{(0)}}_{,L}=-P,\label{6}
\end{equation}	
with $L_{,V}\;{F^{(0)}}_{,L}=-P^{(0)}$ and $P^{(0)}V=\frac{U^{(0)}}{3}$. Hence, from (\ref{14}) we get
\begin{equation}
\hbar\frac{L}{3V}\;{\Phi}_{,L}-\frac{U^{(0)}}{3V}=-P.	
\label{7}
\end{equation}	
After using the (\ref{5}) with $PV=\gamma U(L)$, from (\ref{7}) we obtain the equation (\ref{3}). 
\end{proof}
In this new formulation, the proposition above is more general than the one in \cite{1}, since it is independent on the particular
expression for $\omega^{(0)}$, provided that the initial massless distribution represents a radiation field.
Note that the role of $\Phi$, since it is independent on the temperature
$T$ of the radiation field,
is similar to the one of the fluctuations in solid state physics. However, at this stage of the treatment, no quantum fluctuations are considered. As shown in \cite{1},
by setting $U(L)=E_{ms}=\frac{c^4}{2G}\frac{L^3}{L_A^2}$, from (\ref{3}) we obtain the solution:
\begin{equation}
\hbar\Phi(L)=(1-3\gamma)\frac{c^4}{8G}\frac{L^3}{L_A^2}.
\label{8} 
\end{equation}
With (\ref{8}), for (\ref{4}) we obtain $\gamma>-1$. Hence, exactly the cosmological constant
case is forbidden with $U=E_{ms}$, i.e. by using 
classical setups. In fact, the only possibility with $\gamma=-1$ is that $U^{(0)}=0$, that is possible, with $N\neq0$, only for 
vanishing thermodynamical temperature
$T=0$. In \cite{1} I named the related cosmological constant as the bare non interacting (i.e. non dressed by quantum fluctuations)
cosmological constant $\Gamma$. These reasonings clearly show that, in order to obtain the case with $\gamma=-1$, quantum modifications must be taken into account, and as a consequence Planckian fluctuations come into action.

\section{Planckian fluctuations}

As it is well known, a complete quantum gravity theory is, at present, not at our disposal. The natural arena for
Planckian fluctuations is to consider a quantum spacetime at Planckian scales. There \cite{3,4,5,6,7}, spacetime coordinates
$q^{\mu}$ are satisfying non-trivial physically motivated uncertainty relations (STUR).
These commutation relations \cite{3,4} generate a non commutative $C^{*}$ algebra $\mathcal{\epsilon}$
acting on a generic abstract Hilbert space $\mathcal{H}$. In a Minkowskian background,
spacetime coordinates $\{q^{\mu}\}$ become selfadjoint operators, with the classical Poincaré symmetry 
lifted at a quantum level for the commutation rules $[q^{\mu},q^{\nu}]=\imath\hbar Q^{\mu\nu}$, with $Q^{\mu\nu}$ a covariant tensor 
only under proper Lorentz transformations. The suitable treatment of a non-commutative curved spacetime is much more 
involved. In particular, the implementation of the diffeomorphism covariance at a quantum level represents a formidable task.
As suggested in \cite{4,5}, the following system should be addressed:
\begin{eqnarray}
& & [q^{\mu},q^{\nu}]=\imath\hbar Q^{\mu\nu}(g_{\mu\nu}), \label{9}\\
& & R_{\mu\nu}-\frac{1}{2}R g_{\mu\nu}=T_{\mu\nu}(\psi),\label{10}\\
& & F(\psi)=0, \label{11}
\end{eqnarray}
where  $T_{\mu\nu}$ is supposed to depend on some field $\psi$ with  equation of motion given by (\ref{11}). 
Note that commutators in (\ref{9}) depend on the background metric $g_{\mu\nu}$ that in turn depends on the commutators
among coordinates $\{q^{\mu}\}$: the solution of (\ref{9})-(\ref{11}) represents a formidable task.
As suggested in \cite{4}, we can adopt a semiclassical approximation, where for the metric {\bf g} its classical solution
is considered and instead of
$T_{\mu\nu}$ we have its expectation value on some allowed state $\{\omega\}$ i.e. ${<T_{\mu\nu}>}_{\omega}=\omega(T_{\mu\nu})$.
A first step toward this task has been obtained in \cite{7} for a Friedmann flat spacetime. The semiclassical metric 
can be written in Cartesian coordinates $\{x,y,z\}$ as:
\begin{equation}
ds^2=-c^2d \omega(\hat{t})^2+a(\omega(\hat{t}))^2\left(dx^2+dy^2+dz^2\right),
\label{12}	
\end{equation}
where $\omega(\hat{t})=t$ denotes the expectation value of the time operator $\hat{t}=q^0$ in a given state $\{\omega\}$
and obviously $t$ belongs to the spectrum of $q^0$. Also the spatial coordinates
$\{x,y,z\}$ are elevated to selfadjoint operators $\{q^i\}$. In \cite{7} the STUR are written in terms of $t$ and of the proper spatial variables ${\eta}^i= a(t) x^i$. The STUR in \cite{7} satisfy commutation relations of the form (\ref{9}), where their
exact expression is not relevant in this paper. What is relevant for our purposes is that \cite{1} there exist maximally localizing states $\{s\}$ minimizing the STUR. For such states we have $c\Delta_s t\sim \Delta_s {\eta}^i\sim \Delta_s {\eta}$ and also 
\cite{1} we have $\Delta_s E\Delta_s t\sim\frac{\hbar}{2}$. Moreover, in a de Sitter universe where 
$s(H_{\overline{\Lambda}})=c\sqrt\frac{\overline{\Lambda}}{3}$, we have $\Delta_s L\sim \Delta_s {\eta}^i$ and since
$\Delta_s {\eta}^i\sim L_P$, we have that $\Delta_s E\sim\frac{c\hbar}{2\Delta_s L}$. As a consequence, we can write down
the so physically motivated expression of the generalized Misner-Sharp energy $U_{ms}(L)$:
\begin{equation}
U_{ms}(L)=\frac{c^4}{2G}\frac{L^3}{L_{\Gamma}^2}+\chi\frac{c^4}{2G}\frac{L_P^2}{L},
\label{13}
\end{equation}
where the first term in (\ref{13}) is the classical contribution $E_{ms}$ with $L_{\Gamma}^2=\frac{3}{\Gamma}$
and the second one represents the correction due to
Planckian fluctuations. The constant $\chi$ is a phenomenological one with $\chi\in (0,k]$, with 
$k\sim 1$, and cannot be fixed at a semiclassical level. As we will see in the next section, this new parameter determines the 
decoherence scale $L_D$. Note that in a quantum spacetime $L$ cannot be set to zero thanks to the STUR with 
$L_{min}\sim L_P$. The solution of (\ref{3}) for $\gamma=-1$ is:
\begin{equation}
\hbar\Phi(L)=\frac{c^4}{2G}\frac{L^3}{L_{\Gamma}^2}+\frac{2\chi c^4}{G}
\frac{L_P^2}{L}\ln\left(\frac{L}{L_0}\right),
\label{14}
\end{equation}
where $L_0$ can be set of macroscopic sizes thus assuring that the existence condition (\ref{4}) is satisfied up to such a macroscopic
scale $L_0$. The dressed cosmological constant, that is a consequence of (\ref{13}),
is given by
\begin{equation}
{\overline{\Lambda}}_L=\Gamma+\frac{3\chi L_P^2}{L^4}.
\label{15}
\end{equation}
At this point of the treatment, it should be noticed that the expression (\ref{15}) denotes the effective cosmological constant within 
a sphere of proper radial radius $L$ and thus its value depends on $L$. This could lead to the bad conclusion that we have introduced a
quintessence field, where the value of the effective cosmological constant depends in practice on the dimensions of the observable universe, but this is not the case. If we want to depict a cosmological constant, $\overline{\Lambda}$ must be constant in space and time at a classical level. 
Hence, a correct way to see the formula (\ref{15}) is the following. The metric at the Planck length can be considered rather 
as inhomogeneous, in space and time. If some average procedure is available at some physical scale $L$, 
(\ref{15}) can be considered as an averaged, emergent cosmological constant at the physical scale $L$. At the Planck scale 
$L_P$ the cosmological constant can have a huge value, the one predicted by quantum field theory, by setting 
$k\sim 1$.
However, at scales above the Planck length
the Planckian fluctuations become less relevant. At a scale $L_D$, the decoherence one, such fluctuations become negligible and as a
result a classical de Sitter universe (\ref{2}) emerges with 
\begin{equation}
{\overline{\Lambda}}_{L_D}={\overline{\Lambda}}=\Gamma+\frac{3\chi L_P^2}{L_D^4}.
\label{16}
\end{equation}
This does happen because the decoherence scale is defined as the one where
an absolute minimum for $U_{ms}$ given by (\ref{13}) is obtained. When this minimum is reached at this scale, the cosmological constant is frozen at this value, and at scales $L\geq L_D$ the expression 
(\ref{16}) holds. For the minimum of $U_{ms}$ from (\ref{13}) we have:
\begin{equation}
L_D={\left(\frac{\chi L_P^2}{\Gamma}\right)}^{\frac{1}{4}}.
\label{17}
\end{equation}
From (\ref{17}) and (\ref{16}) we obtain $\overline{\Lambda}(L=L_D)=4\Gamma$. Hence, for $L\in[L_P, L_D)$, the system evolves 
with an energy given by (\ref{13}), while for $L\geq L_D$ the crossover to classicality is reached and the system evolves with
energy $U_{ms}=\frac{c^4 L^3}{2G L_{\overline{\Lambda}}^2}$ with $L_{\overline{\Lambda}}^2=\frac{3}{\overline{\Lambda}}=\frac{3}{4\Gamma}$.
The dressed observed cosmological constant is given by
$\overline{\Lambda}=4\chi\frac{L_P^2}{L_D^4}$, thus depending on the phenomenological quantum 
gravity parameters $\chi$ and $L_D$. For $\chi\sim 1$, we have the observed value for 
$\overline{\Lambda}\sim 10^{-52}/m^2$ for $L_D\sim 10^{-5}m$, implying that classicality is reached at volumes of the order
of $10^{-9} cm^3$. This could be a reasonable value for $L_D$, but unfortunately, I have not  experimental arguments to confirm
this result. For a less value for $\chi$ with $\chi<1$, a lower decoherence scale is obtained that for $\chi<<1$ can be made near the Planck
scale $L_P$. The important fact is that our model furnishes a physical mechanism, formula (\ref{3}), transforming a given radiation
field, thanks to quantum fluctuations, to one with the correct equation of state and also is capable to explain the smallness of the 
cosmological constant.\\
In our treatment we used the model in \cite{7} to obtain the expression (\ref{13}) representing the generalized Misner-Sharp
energy. However, to go further we need to depict our idea dynamically, by solving the equation (\ref{10}) in a quantum spacetime.
Unfortunately, this task is at present day unattainable since a physically sound quantum gravity theory is not at our disposal.
In order to study the birth of the cosmological constant dynamically, we can use once again a semiclassical approximation
by a suitable modification and study of the well known Buchert formalism \cite{2}: this will be done in the next section.

\section{A semiclassical solution: averaging quantum fluctuations}

\subsection{Fitting problem: from cosmological to Planckian scales}

As it is well known, our universe on small scales is very lumpy, with clusters and superclusters of galaxies forming a kind of
web structure. However, when bigger and bigger scales are considered, our universe, on average, becomes practically homogeneous and isotropic. As a consequence our universe, according to astrophysical data, is very well depicted in terms of a
Friedmann flat metric equipped with a small non-vanishing cosmological constant. Nevertheless, it is interesting to 
study the effects of these small scale inhomogeneities on the cosmological parameters. To this purpose, 
the Buchert equations \cite{2} have been remarkably used in literature in order to study the effects of the inhomogeneities present at 
small cosmological scales on the evolution of the universe \footnote{See also the paper \cite{24} and
references therein.}. In particular, Buchert equations \cite{2} are an attempt to solve the fitting problem: a given inhomogeneous metric, representing an exact solution of Einstein equations for a real universe where inhomogeneities are present,
is interpreted in terms of a constant curvature hypersurface by means of a template metric, provided that backreaction 
$\mathcal{Q}$ is taken into account by means of Buchert equations.
This template metric is not an exact solution of Einstein equations, but rather of the Buchert ones. Kinematical
backreaction is a consequence of the smoothing out procedure of the inhomogeneities. The backreaction has been also debated in the literature (see for example \cite{25}) as a possible source for the cosmological constant, but its effect seems to be completely insufficient in order to explain the actual value of $\overline{\Lambda}$. This is because kinematical backreaction
becomes practically negligible when averaged at the scale of homogeneity $L_o$, i.e.
$\mathcal{Q}_{L_o}\simeq 0$. Remember that the vanishing of $\mathcal{Q}$ is a necessary and sufficient condition in order
to regain the classical Einstein equations.\\
The Buchert scheme has been proposed in \cite{1} in order to study in a semiclassical approximation the
evolution of a background metric in a quantum spacetime. In practice, the lumpy universe at small scales is translated at 
Planckian scales in order to depict, in a semiclassical way, the strong fluctuations at Planckian scales.
The scale of homogeneity $L_o$ at a cosmological context is translated into the $L_D$ one, denoting the crossover to 
classicality. We expect a very large class of initial conditions at $t=t_0$ such that 
$\mathcal{Q}_{L_D}(t_0)\simeq 0$, ensuring the transition to classicality with a classical de Sitter spacetime.
Also in \cite{17} a 
preliminary study of an emergent cosmological constant from Planckian fluctuations is present. There, attention is posed on the initial conditions assuring that the initial spatial curvature becomes smaller and smaller when averaged 
\footnote{Also in \cite{17} a Buchert scheme is proposed in order to study Planckian fluctuations.} 
on bigger and bigger scales
with respect to the Planck scale.
However, the model in \cite{17} cannot explain the smallness of $\overline{\Lambda}$, and a study of the evolution of the initial conditions, due to the dependence on the time coordinate chosen,  is not a simple task. In \cite{1} 
I have proposed to use the Buchert scheme where the time coordinate $t$ belongs to the spectrum of the time operator $\hat{t}$, together with the use of 
spherical maximally localizing states saturing the STUR \cite{3,4}.\\
To be more concrete, the starting point, in line with the semiclassical approximation discussed above
\footnote{See also the interesting paper in \cite{15}, where a similar semiclassical approximation has been studied in terms 
of an inhomogeneous metric at Planckian scales.}, is the inhomogeneous metric at Planckian scales given by:
\begin{equation}
ds^2=-N(t, x^i)c^2 dt^2+h_{ij}(t, x^i)dx^i dx^j,
\label{18}
\end{equation}
where $N(t, x^i)$ denotes the lapse function, and $\{t, x^i\}\in\{sp(\hat{t}, sp({\hat{x}}^i))\}$, with $\{x^i\}$ the usual
Cartesian coordinates with $\hat{x^i}$ denoting quantum operators acting on some abstract Hilbert space $\mathcal{H}$.
Obviously, the coordinates in the semiclassical metric (\ref{18}) can be seen as expectation values on some state $\{\omega\}$.
In the following we use privileged states that are, as explained in the  section above, maximally localizing states
$\{s\}$ saturing the STUR \cite{7}: $\{t, x^i\}=\{s(\hat{t}), s({\hat{x}}^i\})$.\\
The further step is to fit in a suitable way the metric (\ref{18}) on a given time slicing of the spacetime 
by means of a template metric, mimicking the cosmological case.
First of all we must fix the hypersurface $\mathcal{S}$ where the spatial average is performed: this is provided by the surfaces
$s(\hat{t})=k\in R$. Hence, according to the Buchert procedure \cite{2}, we can consider a given proper spherical volume 
$V(\mathcal{S})$ given by
\begin{equation}
V_{\mathcal{S}}=\int_{\mathcal{S}}\sqrt{g^{(3)}}d^3x, 
\label{19}
\end{equation}
where  $g^{(3)}$ denotes the determinant of the three-metric $h_{ij}$ on the slice $\mathcal{S}$. For any scalar field
$\psi(s(t), s(x^i))$ its spatial average is:
\begin{equation}
{<\psi(s(\hat{t}), s({\hat{x}}^i)>}_{V_{\mathcal{S}}}=\frac{1}{V_{\mathcal{S}}}\int_{\mathcal{S}}
\psi(s(\hat{t}), s({\hat{x}}^i))\sqrt{g^{(3)}}d^3x,
\label{20}
\end{equation}
For the averaged expansion parameter ${<\theta>}_{\mathcal{S}}$ and the effective volume scale factor 
$a_{\mathcal{S}}(s(\hat{t}))$ we have:
\begin{equation}
{<\theta>}_{V_{\mathcal{S}}}=\frac{{\dot{V}}_{\mathcal{S}}}{V_{\mathcal{S}}}=
3\frac{{\dot{a}}_{V_{\mathcal{S}}}}{a_{V_{\mathcal{S}}}},\;\;
a_{V_{\mathcal{S}}}(s(\hat{t}))={\left(\frac{V_{\mathcal{S}}(s(\hat{t}))}{V_{\mathcal{S}}(s({\hat{t}}_0))}\right)}^{\frac{1}{3}}
\label{21}
\end{equation}
In order to address the fitting problem on Planckian scales, one must analyse the lapse function in (\ref{18}).
First of all note that, thanks to the huge inhomogeneities at Planckian scales, the lapse function is expected to vary also in space. The choice $N=1$ can be used if the gradient in the space of the
gravitational energy can be neglected \cite{24}. As an example and in a cosmological context, in \cite{24} the lapse function is averaged to unity at the scale of homogeneity $L_o$
($L_o\sim 70-100 Mpc$), while it
is expected to be different from the unity on smaller scale below $L_o$ and in
presence of a non-vanishing spatial curvature. It is reasonable to expect that
a similar phenomenon does appear at microscopic scales, where the role of the cosmological scale $L_o$ is played in our context
by the decoherence scale $L_D$, given by (\ref{17}) and representing the absolute minimum (true vacuum state) for the generalized 
Misner-Sharp energy $U_{ms}$ in (\ref{13}). 
Moreover, according to our translation of the cosmological context at Planckian scales,
a relation between $N$ and the behavior of the spatial curvature $^{(3)}R$ emerges: for huge spatial inhomogeneities
we expect  values for $N$ different from the unity. In particular, for the averaged value of 
$N$, i.e. ${<N(s(\hat{t}, s({\hat{x}}^i)))>}_{\mathcal{S}}=\mathcal{N}(s(\hat{t}))$, we generally expect values greater than one
with negative  averaged spatial curvatures (positive gravitational energy), and values less than one for positive 
averaged spatial curvatures
(negative gravitational energy). As shown in \cite{17}, although some physical arguments hint a small value for the averaged
spatial curvature
$\mathcal{R}\simeq 0$, also by assuming a non-negligible spatial curvature at Planckian scales, there exists a large class of initial 
conditions assuring decaying spatial curvature when averaged at small macroscopic scales. Hence,  when the averaging
procedure is outlined on bigger and bigger scales with $L_D>L_{I_1}>L_{I_2}>\cdots>L_P$ we expect, for averaged negative spatial
curvatures $\mathcal{R}<0$, that\footnote{See \cite{25} in a cosmological context.}:
\begin{equation}
1={\mathcal{N}}_{L_D}\;<\;{\mathcal{N}}_{L_{I_1}}\;<\;{\mathcal{N}}_{L_{I_2}}\;<\;\cdots\;<{\mathcal{N}}_{L_{P}},
\label{22}
\end{equation}
while for averaged positive spatial curvatures $\mathcal{R}>0$ we must have:
\begin{equation}
1={\mathcal{N}}_{L_D}\;>\;{\mathcal{N}}_{L_{I_1}}\;>\;{\mathcal{N}}_{L_{I_2}}\;>\;\cdots\;>{\mathcal{N}}_{L_{P}}.
\label{23}
\end{equation}
As a consequence, if a non-vanishing non-negligible statial curvature $\mathcal{R}$ at Planckian scales is assumed, we expect a
decreasing $\mathcal{R}$ when averaged on bigger scales than the Planck one and at the decoherence scale $L_D$ we can assume
that ${\mathcal{N}}_{L_D}=1$.\\
Finally, for the case of negligible spatial curvature also at Planckian scales, we are legitimate to assume $\mathcal{N}=1$
also at scales below the decoherence scale.\\
For the reasonings above, we separately study the dynamics of our semiclassical model for $\mathcal{R}=0$
and $\mathcal{R}\neq 0$. In all the cases discussed, the effective cosmological constant is provided by (\ref{15}) with
the energy $U_{ms}$ given by (\ref{13}).

\subsection{Dynamical evolution with $\mathcal{R}=0$}
For a vanishing spatial curvature at scales $L$ the effective template metric,
after using the short notation
$t=s(\hat{t}), x^i=s({\hat{x}}^i)$, can be written as:
\begin{equation}
ds^2=-c^2 dt^2+a_{L}^2(t)\left[dr^2+r^2\left(d\theta^2+\sin^2\theta d\phi^2\right)\right],
\label{24} 
\end{equation}
where $a_L(t)$ denotes the effective volume scale factor in (\ref{21}) at the scale $L$ with $L$ the areal
radius of the spatial average surface $\mathcal{S}$. The relevant generalized Buchert equations at 
microscopic Planckian scales for irrotational fluids are:
\begin{eqnarray}
& &3\frac{{\dot{a_L}}^2}{a_L^2}=
c^2\Gamma+\frac{3c^2\chi L_P^2}{L^4}
-\frac{\mathcal{Q}_L}{2}=c^2{\overline{\Lambda}}_L-\frac{\mathcal{Q}_L}{2},\label{25}\\
& & 6\mathcal{Q}_L\dot{a_L}+a_L\dot{\mathcal{Q}_L}=0, \label{26}\\
& &\mathcal{Q}_L=\frac{2}{3}\left[{<{\theta}^2>}_{L}-{<\theta>}^2_L\right]-2{<\sigma>}^2_L,\label{27}
\end{eqnarray}
where ${\mathcal{Q}}_L$ is the kinematical backreaction at the scale $L$ 
and $\sigma$ denotes the shear and dot denotes time derivative with
respect to $t$. 
Similarly to the average problem in a cosmological context
\cite{24}, the shear parameter $\sigma$ is expected to be more and more smaller when the decoherence scale is approached and it is expected to be 
non-negligible near the Planck scale. On general grounds we can suppose, as happens in a cosmological context \cite{24}, that ${\mathcal{Q}}_L>0$ on sufficiently large 
microscopic scales, while for $L\sim L_P$, the variance in the shear can be very large and
the situation ${\mathcal{Q}}_L<0$ can arise. Hence, in the following  we suppose a positive backreaction. The case of negative backreaction will be analysed at subsection $4.4$.
To integrate the equation (\ref{25})), one must first integrate the
equation (\ref{26}):
\begin{equation}
\mathcal{Q}_L(t)=\mathcal{Q}_L(t_0){\left(\frac{a_L(t_0)}{a_L(t)}\right)}^6.
\label{28}
\end{equation}
From (\ref{28}) it is evident that, if the system evolves with an averaged
scale factor $a_L(t)$ monotonically increasing  with respect to $t$, then the kinematical backreaction becomes vanishing at late times. To confirm this fact,
we put the solution (\ref{28}) in (\ref{25}). We obtain:
\begin{equation}
\frac{{\dot{a_L}}^2}{a_L^2}=c^2\frac{{\overline{\Lambda}}_L}{3}-\frac{\mathcal{Q}_L(t_0)}{6}{\left(\frac{a_L(t_0)}{a_L(t)}\right)}^6.
\label{29}
\end{equation}
Equation (\ref{29}) can be integrated to obtain:
\begin{eqnarray}
& & c {a_L^3}(t)\sqrt{\frac{{\overline{\Lambda}}_L}{3}}+
\sqrt{c^2 a_L^6(t)\frac{{\overline{\Lambda}}_L}{3}-\frac{\mathcal{Q}_L(t_0)}{6}a_L^6(t_0)}\;=\nonumber\\
& & =K_0 e^{3c(t-t_0)\sqrt{\frac{{\overline{\Lambda}}_L}{3}}},
\label{30}
\end{eqnarray}
where
\begin{equation}
K_0 = c {a_L^3}(t_0)\sqrt{\frac{{\overline{\Lambda}}_L}{3}}+
\sqrt{c^2 a_L^6(t_0)\frac{{\overline{\Lambda}}_L}{3}-\frac{\mathcal{Q}_L(t_0)}{6}a_L^6(t_0)}.
\label{31}
\end{equation}
From (\ref{31}) the following existence condition holds:
\begin{equation}
\mathcal{Q}_L(t_0)\leq 2c^2 {\overline{\Lambda}}_L,
\label{32}
\end{equation}
with the conditon $\mathcal{Q}_L(t_0)\geq 0$ discussed above.\\
Condition (\ref{32}) is important because it implies that if initial conditions are such that the kinematical backreaction 
violates this inequality, we have no solution to the fitting problem. 
Note that a huge value for $\mathcal{Q}_L(t_0)$ means huge inhomogeneities for the metric (\ref{18}). 
This can be also interpreted in light of the results in
\cite{18} and cited in \cite{17}:
a spacetime, solution of Einstein equations, equipped with a 
positive cosmological constant but with 
sufficiently strong
inhomogeneities does not evolve and as a consequence a de Sitter spacetime cannot emerge. 
Translated into the Buchert language, if (\ref{32}) is violated, backreaction is not longer negligible and the resulting 
region evolves differently from the behavior predicted by the Friedmann paradigm. 
In this case, Planckian fluctuations are expected to be rather huge, and the transition to classicality,
i.e. to a classical de Sitter universe, is practically absent.\\
With the condition (\ref{32}), the explicit solution  for (\ref{30}) is:
\begin{equation}
a_L(t)=\frac{1}{{\left[2K_0 c\sqrt{\frac{{\overline{\Lambda}}_L}{3}}\right]}^{\frac{1}{3}}}
{\left[K_0^2\;e^{3c(t-t_0)\sqrt{\frac{{\overline{\Lambda}}_L}{3}}}+
	\frac{\mathcal{Q}_L(t_0) a_L^6(t_0)}{6 e^{3c(t-t_0)\sqrt{\frac{{\overline{\Lambda}}_L}{3}}}}\right]}^{\frac{1}{3}}.
\label{33}
\end{equation}
As evident from (\ref{33}), a de Sitter universe solution of the classical Einstein equations emerges only at times
for $t\rightarrow\infty$.
At the decoherence scale $L_D$, the spacetime 
has a minimum in the energy $U_{ms}$ and as a consequence the cosmological constant remains frozen at the value given by
(\ref{16}). Concerning the expression for $\mathcal{Q}_L(t_0)$, it is expected to be
a monotonically decreasing function of the physical length scale where averaging is performed. In particular, it is expected that,
at the decoherence scale $L_D$, with (\ref{32}), one has
$\mathcal{Q}_{L_D}(t_0)\simeq 0$. As a result, at the scale $L_D$ the spacetime exactly becomes the 
classical one (\ref{2}) solution of Einstein equations with the frozen 
observed cosmological constant $\overline{\Lambda}$. Moreover, at the decoherence scale we have the 
crossover to classicality and the mean values $\{s(\hat{t}), s({\hat{x}}^i)\}$ behave as the classical coordinates and thus 
a classical de Sitter universe emerges with a small frozen cosmological constant. As a final consideration, if for some model we
relax the physically reasonable
hypothesis $\mathcal{Q}_L>0$ and consider the one with $\mathcal{Q}_L<0$, we have no restriction from 
(\ref{31}).

\subsection{Dynamical evolution with $\mathcal{R}\neq 0$: a power law behavior}

The case of an initial non-vanishing spatial curvature ${\mathcal{R}}_{L_D}(t_0)$ is more complicated than the one with 
$\mathcal{R}=0$. To start with, the relevant equations for irrotational fluids at the decoherence scale $L_D$ are:
\begin{eqnarray}
& &3\frac{{\dot{a}}_{L_D}^2}{a_{L_D}^2}=\nonumber\\
& & =c^2\Gamma+\frac{3c^2\chi L_P^2}{L_D^4}
-\frac{\mathcal{Q}_{L_D}}{2}-\frac{{\mathcal{R}}_{L_D}}{2}=c^2{\overline{\Lambda}}-\frac{\mathcal{Q}_{L_D}}{2}-
 \frac{{\mathcal{R}}_{L_D}}{2},\label{34}\\
& & {\partial}_t\left(a_{L_D}^6 \mathcal{Q}_{L_D}\right)+
 {\partial}_t\left(a_{L_D}^6 \mathcal{R}_{L_D}\right)=
 a_{L_D}^2\mathcal{R}_{L_D} {\partial}_t\left(a_{L_D}^4\right), \label{35}
\end{eqnarray}
where ${\mathcal{Q}}_{L_D}$ is given by (\ref{27}) with obviously $L\rightarrow L_D$. The first step is the integration of 
(\ref{35}). Obviously this equation cannot be integrated directly because it does not represent an exact differential. 
First of all, note that (\ref{35}) can be rewritten as:
\begin{equation}
 {\partial}_t\left(a_{L_D}^6 \mathcal{Q}_{L_D}+a_{L_D}^6 \mathcal{R}_{L_D}\right)=a_{L_D}^2 \mathcal{R}_{L_D}
 {\partial}_t\left(a_{L_D}^4\right)
\label{36}
\end{equation}
From a first inspection of (\ref{36}), by setting $\mathcal{Q}_{L_D}+\mathcal{R}_{L_D}=0$, from (\ref{34}) we realize that the
only possible solution is $\mathcal{Q}_{L_D}=\mathcal{R}_{L_D}=0$, the trivial solution leading to the classical Friedmann
equations. Another simple possibility is provided by the particular solution:
\begin{equation}
\mathcal{Q}_{L_D}+\mathcal{R}_{L_D}=\frac{c_0}{a_{L_P}^6}.
\label{37}
\end{equation}
Under the assumption that for the constant
$c_0$ we have $c_0>0$, from (\ref{37}) with $\mathcal{R}_{L_D}\neq 0$, we have $a_{L_D}=q\in R^+$, with the condition
from (\ref{34}) $c^2{\overline{\Lambda}}=\frac{c_0}{2q}$: in practice, with this particular solution, backreaction due to a 
non-vanishing spatial curvature generates an averaged spacetime with an effective vanishing cosmological constant and thus the 
resulting spacetime is stationary on average. Apart from these particular fine tuned
solutions, we expect, for a large class of initial conditions,
as shown in \cite{17}, an expanding spacetime with an averaged spatial curvature negligible at the decoherence scale.
It is in fact
physically reasonable  to impose that the averaged quantities $\mathcal{Q}_{L_D}$ and $\mathcal{R}_{L_D}$ are monotonically decreasing functions
of $L_D$ and $t$: $\mathcal{Q}_{L_D}=\mathcal{Q}_{L_D}(t,L_D),\;\mathcal{R}_{L_D}=\mathcal{R}_{L_D}(t, L_D)$. We could in principle invert these relations to obtain $t=H(\mathcal{Q}_{L_D}, \mathcal{R}_{L_D})$ and 
$L_D=K(\mathcal{Q}_{L_D}, \mathcal{R}_{L_D})$. Hence, on general grounds we have a functional relation between 
$\mathcal{Q}_{L_D}$ and $\mathcal{R}_{L_D}$: $g(\mathcal{Q}_{L_D}, \mathcal{R}_{L_D})=0$. Since a functional relation is chosen, equation (\ref{36}) can be integrated. Since, as stated above, we expect that for $\mathcal{Q}_{L_D}$ small also
$\mathcal{R}_{L_D}$ is small, the simplest and reasonable assumption, also for practical purposes, is 
\begin{equation}
\mathcal{Q}_{L_D}=\alpha \mathcal{R}_{L_D}, \;\;\;\alpha\in R. 
\label{38}
\end{equation} 
Also in this section, we retain the condition $\mathcal{Q}_{L_D}>0$. 
With (\ref{38}), the solution of (\ref{36}), with $\alpha\neq -1$, is given by:
\begin{equation}
\mathcal{R}_{L_D}=\mathcal{R}_{L_D}(t_0){\left(\frac{a_{L_D}(t_0)}{a_{L_D}(t)}\right)}^{\frac{2(1+3\alpha)}{(1+\alpha)}}.
\label{39}
\end{equation}
From (\ref{39}) we deduce that, for an expanding averaged spacetime, backreaction is decreasing in time if and only if
$\alpha\in (-\infty, -1)\cup \left(-\frac{1}{3}, +\infty\right)$. For the setups above, for negative values for $\alpha$ we have
$\mathcal{R}_{L_D}<0$ while for positive values we have $ \mathcal{R}_{L_D}>0$. With (\ref{39}), equation
(\ref{34}) can be integrated to obtain:
\begin{equation}
a_{L_D}(t)=
\frac{1}{{\left[2K_{1} c\sqrt{\frac{{\overline{\Lambda}}}{3}}\right]}^{\frac{1+\alpha}{1+3\alpha}}}
{\left[K_1^2\;e^{\frac{1+3\alpha}{1+\alpha}c(t-t_0)\sqrt{\frac{{\overline{\Lambda}}}{3}}}+
\frac{(1+\alpha)\mathcal{R}_{L_D}(t_0) a_{L_D}^{\frac{2(1+3\alpha)}{(1+\alpha)}}(t_0)}{6 e^{\frac{1+3\alpha}{1+\alpha}c(t-t_0)\sqrt{\frac{{\overline{\Lambda}}}{3}}}}\right]}^{\frac{1+\alpha}{1+3\alpha}},
\label{40}
\end{equation}
where
\begin{equation}
K_1 = c {a_{L_D}^{\frac{1+3\alpha}{1+\alpha}}}(t_0)\sqrt{\frac{{\overline{\Lambda}}}{3}}+
a_{L_D}^{\frac{1+3\alpha}{1+\alpha}}(t_0)\sqrt{c^2\frac{{\overline{\Lambda}}}{3}-\frac{(1+\alpha)\mathcal{R}_{L_D}(t_0)}{6}}.
\label{41}
\end{equation}
Similarly to equation (\ref{32}), from (\ref{41}) we deduce the existence condition:
\begin{equation}
2c^2{\overline{\Lambda}}\geq (1+\alpha)\mathcal{R}_{L_D}(t_0).
\label{42}
\end{equation}
Condition (\ref{42}) means that, if the initial conditions are such that the initial averaged spatial curvature
$\mathcal{R}_{L_D}$ is sufficiently small with $\mathcal{R}_{L_D}\simeq 0$ ($\mathcal{Q}_{L_D}\simeq 0$), 
then we have the crossover to classicality and
(\ref{40}) reduces to the exact solution (\ref{2}).
This simple instructive example and the study of the case $\mathcal{R}(t_0)=0$ 
show that there exists, as shown in \cite{17}, a large class of
initial conditions leading to an emergent de Sitter spacetime with a small cosmological constant, thus giving a physically
reasonable solution to the cosmological constant problem, without introducing quintessence fields.\\
As noticed \footnote{Note that in \cite{26} these solutions have been studied in a dust filled universe with a vanishing
cosmological constant in order to explain $\overline{\Lambda}$ in terms of $\mathcal{Q}$.}
in \cite{26}, if we look for solutions of the integrability condition (\ref{35}) in terms of power law
behavior:
\begin{equation}
\mathcal{Q}_L\sim a_L^p,\;\;\;\;\mathcal{R}_L\sim a_L^n,
\label{b1}
\end{equation}
only two cases are possible. The first one is with $n=p$, that is the one considered in this section by (\ref{38}).
This solution corresponds to the case where there exists a direct strong coupling between $\mathcal{Q}$ and
$\mathcal{R}$, and is the case where kinematical backreaction is expected to be huge. The other solution is the one with
$n=-2,\;p=-6$ and it is expected to be suitable for the cases with negligible backreaction. Also in this case the condition
$\mathcal{Q}_{L_D}(t_0)=0$ is a necessary and sufficient condition to ensure the crossover to classicality.

\subsection{Morphon field, initial conditions and caveats}

In order to study more general solutions, it is helpful to introduce the 'morphon' field defined in \cite{26} in terms of
a scalar field $\varpi_{L}$ and evolving in an effective potential $U_L$:
\begin{eqnarray}
& & -\frac{1}{8\pi G}\mathcal{Q}_L=\epsilon{\dot{\varpi}}_L^2-U_L,\label{b2}\\
& &-\frac{1}{8\pi G}\mathcal{R}_L = 3U_L, \label{b3}
\end{eqnarray}
where \cite{26} $\epsilon=+1$ for a field with a positive kinetic energy and $\epsilon=-1$ with a negative kinetic energy.
The field $\varpi_{L}$ satisfies a Klein-Gordon equation
\begin{equation}
{\ddot{\varpi}}_L+{<\theta>}_L {\dot{\varpi}}_L+\epsilon{U_{L,{\varpi_{L}}}}(\varpi_{L})=0,
\label{b4}
\end{equation}
where comma denotes partial derivative.
The kinematical backreaction can thus be written as:
\begin{equation}
\mathcal{Q}_L=-\frac{\mathcal{R}_L}{3}-\epsilon 8\pi G {\dot{\varpi}}_L^2.
\label{b5}
\end{equation}
In relation to the general case with $\mathcal{R}_L=0$, the necessary and sufficient initial condition 
for the birth of the de Sitter universe at
$L=L_D$ is the following:
\begin{equation}
 {\dot{\varpi}}_{L_D}(t_0)=0\;\rightarrow {\varpi_{L_D}}(t_0) = k\in\Re. 
\label{b6}
\end{equation}
The (\ref{b6}) is similar to the slow roll inflation condition, but applied to the scalar field 
$\varpi_{L}$. In the case studied in the subsection $4.3$ with $\mathcal{Q}_{L_D}=\alpha \mathcal{R}_{L_D}$
we have:
\begin{equation}
\epsilon {\dot{\varpi}}_{L_D}^2=\left(-\alpha-\frac{1}{3}\right)\frac{\mathcal{R}_{L_D}}{8\pi G}
\label{b7}
\end{equation}
In order to have a time-decreasing expression for $\mathcal{R}_{L_D}$, equation (\ref{39}), we must impose
$\alpha\neq -\frac{1}{3}$. Hence, initial condition (\ref{b6}) still holds. In more general cases,
we may think of a complicated relation between $\mathcal{Q}_{L_D}$ and $\mathcal{R}_{L_D}$.
In such a case, condition (\ref{b6}) is necessary but not sufficient.\\ 
However, it is reasonable that spatial curvature, created 
by fluctuations, is strongly coupled to kinematical backreation in such a way that 
$\mathcal{Q}_{L_D}(t_0)=\mathcal{R}_{L_D}(t_0)=0$. As a result, the following condition becomes necessary
and sufficient:
\begin{equation}
{\dot{\varpi}}_{L_D}(t_0)=0,\;\;\;\;U_{L_D}({\varpi_{L_D}}(t_0))=0.
\label{b8}
\end{equation}
Hence, the condition for the crossover to classicality is provided by a kind of slow-roll initial condition for the scalar field
$\varpi_{L_D}$, created by vacuum fluctuations.\\
For a first look at case 
with negative backreaction, consider the Buchert equation
\footnote{This equation is dependent on the system (\ref{34})-(\ref{35})}:
\begin{equation}
3\frac{{\ddot{a}}_{L_D}}{a_{L_D}}=\overline{\Lambda}+\mathcal{Q}_{L_D}.
\label{b9}
\end{equation}
A positive backreaction mimicks a positive cosmological constant, while a negative backreaction struggles with
$\overline{\Lambda}$. In the case with $\mathcal{Q}_{L_D}<0$, we have from (\ref{b9}) an accelerating
(${{\ddot{a}}_{L_D}}>0$)
template spacetime for
$\overline{\Lambda} > -\mathcal{Q}_{L_D}$, while a decelerating one is obtained for
$\overline{\Lambda} < -\mathcal{Q}_{L_D}$. Concerning the existence condition, in the case with 
$\mathcal{R}_{L_D}=0$, it is satisfied for a positive $\overline{\Lambda}$. Moreover, with $\mathcal{R}_{L_D}\neq 0$,
we have the condition:
\begin{equation}
c^2{\overline{\Lambda}}-\frac{\mathcal{Q}_{L_D}}{2}-
\frac{{\mathcal{R}}_{L_D}}{2}>0.
\label{b10}
\end{equation}
Condition (\ref{42}) is a particular case of (\ref{b10}). For realistic models with time-decreasing 
$|\mathcal{Q}_{L_D}|$ and $|\mathcal{R}_{L_D}|$, existence condition follows from equation 
(\ref{34}):
\begin{equation}
c^2{\overline{\Lambda}}-\frac{\mathcal{Q}_{L_D}(t_0)}{2}-
\frac{{\mathcal{R}}_{L_D}(t_0)}{2}>0.
\label{b11}
\end{equation}
Note that (\ref{b11}) is a general condition, independent on the sign of $\mathcal{Q}_{L_D}$. For
$\mathcal{Q}_{L_D}(t_0)<0$ the (\ref{b11}) is non-trivial only when
$\mathcal{R}_{L_D}(t_0)>0$. This is because a positive curvature struggles with the cosmological constant
$\overline{\Lambda}$. Otherwise, for $\mathcal{Q}_{L_D}(t_0)>0$ (\ref{b11}) is always non-trivial.\\
If (\ref{b11}) is violated, no solution to the fitting problem is present. This result, once again, 
can be interpreted in light of the results in \cite{18}: if initial conditions are such that inhomogeneities are sufficiently 
strong, the $\overline{\Lambda}$ is hidden.\\
As a final consideration, some reasonings about the limitations of the Buchert-like approach proposed in this paper can be 
outlined.
At cosmological scales, the Buchert equations give a solution to the fitting problem by means of a template metric
and initial conditions are given on a constant curvature hypersurface. This fitted metric is not a solution of the exact Einstein equations. As a consequence, 
it is far from obvious that a given initial condition imposed on 
$\mathcal{Q}_{L_D}$ and  $\mathcal{R}_{L_D}$ is compatible with an inhomogeneous spacetime (\ref{18}) that in turn is a solution
of the Einstein equations. The same obviously generally applies to our 'microscopic' translation of the Buchert formalism.
The metric (\ref{18}) is expected to satisfy Einstein's equations but with a 
semiclassical approximation, i.e. $G_{\mu\nu}=\frac{8\pi G}{c^4}{<T_{\mu\nu}>}_s$ for some state $\{s\}$
with the expression for 
$T_{\mu\nu}=-8\pi G{\rho}_{vac} g_{\mu\nu}$. After performing the average at the proper volume (\ref{19}) at the scale
$L$, we have the template metric $a_L$ together with the effective cosmological constant ${\overline{\Lambda}}_L$, where
the observed value is fixed at the scale $L_D$. However, a difference from the cosmological case arises. In
a cosmological context, backreaction has been invoked \cite{26} to explain the acceleration of the universe
without $\overline{\Lambda}$. To this purpose, a huge backreaction, strongly 
coupled with a huge negative curvature, is required 
also at scales well above the homogeneity one ($\sim 70$ Mpc). It is thus far from obvious that \cite{26} the averaged spacetime
can be obtained starting from a realistic metric (\ref{18}). It can be stressed that, differently from the model in \cite{26},
the model of this paper does not depict the birth of $\overline{\Lambda}$ in terms of $\mathcal{Q}_{L_D}$, but instead, in order that the classical solution (\ref{2}) emerges above $L_D$, initial condition (\ref{b8}) is imposed at $t=t_0$. 
As a consequence, it is not required to have a particular expression for $\mathcal{Q}_{L_D}$ and $\mathcal{R}_{L_D}$, but only
the vanishing of these quantities, expressed in (\ref{b8}) in terms of the useful 'morphon' formalism, at $t=t_0$.
It is thus expected, also according to the finding in \cite{17}, that there exists a wide range of realistic models with
(\ref{b8}) satisfied, with the expression for $\mathcal{Q}_{L_D}$ depending, via averaging procedure, on the 'exact' metric
(\ref{18}) that in turn is a consequence of the Planckian model used to depict quantum fluctuations of the 
spacetime geometry. This can certainly matter for a future investigation with an explicit estimation concerning the 
'phenomenological' parameters $\chi$ and $L_D$.

\section{Conclusions and final remarks}

In this paper I have continued the study presented in \cite{1} concerning the birth and features of the cosmological constant.
In my view, the cosmological constant is generated at Planckian scales, where quantum fluctuations come into action and are supposed to be very strong. To this purpose, a mechanism mimicking the solid state physics \cite{1}
is proposed transforming, thanks to these huge Planckian fluctuations, a radiation 
field into one with the cosmological constant equations of state. The usual way to look at vacuum energy, where contributions arise from different physical scales, could be incorrect since these contributions 
represent radiation fields and thus they cannot contribute to the cosmological constant: only for a radiation field sufficiently near 
the Planck scale $L_P$ fluctuations are so strong to permit its transformation into a cosmological constant equation of state,
via equation (\ref{3}).
A further ingredient is provided by noticing that energy contributions enter in Einstein's equations in terms of energy-densities and thus one must specify the physical scale such that a given contribution is averaged. To this purpose, I propose that the cosmological constant born at Planckian scales acts also at bigger scales, but with a monotonically decreasing behavior in terms of the length scale $L$, namely equation (\ref{15}), motivated by a non-commutative spacetime. Since the modified Misner-Sharp energy (\ref{13}) has an absolute minimum at $L=L_D$
given by (\ref{17}), this minimum represents the decoherence scale where the crossover at the classicality is recovered:
this scale determines the observed small value of the cosmological constant $\overline{\Lambda}$. This picture is different from the interesting proposal in \cite{17}. There, the cosmological constant is also believed to be generated at Planckian scales 
but it is also assumed that this cosmological constant is hidden by the Planckian fluctuations and as a result the spacetime is expanding with a residual cosmological constant at macroscopic scales. This picture, although interesting, does not explain the smallness of $\overline{\Lambda}$. Moreover, the paper in \cite{17} contains the important information regarding the existence
of a large class of initial conditions leading to a negligible spatial curvature at macroscopic scales.\\ 
In this paper I have
depicted dynamically the new scenario in \cite{1}. To this purpose,
one should solve the system (\ref{9})-(\ref{11}) proposed in \cite{4,5}; a formidable task. To make the problem more tractable, 
a semiclassical approximation has been used. In a semiclassical approximation context,
the starting point is the inhomogeneous metric (\ref{18}), where for the coordinates we have
$\{t, x^i\}\in\{sp(\hat{t}, sp({\hat{x}}^i))\}$. In particular, privileged states exist, namely maximally localizing states 
$\{s\}$ \cite{4,5} saturing the STUR of a quantum spacetime \cite{7}. With the semiclassical metric (\ref{18}), the next step is to 
introduce an averaging procedure, mimicking the Buchert formalism \cite{1} but applied in order to average out the Planckian 
fluctuations. This procedure can be obtained with an averaging procedure on the slice at $s(\hat{t})=constant.$ In this paper I 
considered separately the case with an initial vanishing spatial curvature from the one with a non-vanishing spatial curvature. In
the former case, the lapse function can be averaged to unity and the average procedure can be outlined also below the 
decoherence scale $L_D$. In the latter case the average procedure can be consistently obtained at the decoherence scale.
In my view, the decoherence scale plays the role of the homogeneity scale $L_o$ in cosmology \cite{24}. What is physically relevant is that, in both cases, we have existence conditions, namely conditions (\ref{32}) and (\ref{42}). These conditions do imply
that the initial backreaction and the initial spatial curvature must be sufficiently small, otherwise the system is not evolving and the fitting problem has no solution. Physically, this is in agreement with the fact
that for an initially very inhomogeneous metric (\ref{18}) solution of Einstein equations, a spacetime 
also equipped with a positive cosmological constant is not evolving as a de Sitter spacetime.
Conversely, according to \cite{17}, there exists a large class of initial conditions assuring a vanishing 
kinematical backreaction and spatial curvature at the scale $L_D$. These initial conditions,
assuring the emergence of a classical
de Sitter universe with a small cosmological constant $\overline{\Lambda}$,
have been expressed in
terms of a morphon field by means of (\ref{b8}).\\
The proposal in \cite{1} that is further studied in this paper, is capable of depicting the birth of the cosmological constant 
with a simple physical process where quantum fluctuations transform a radiation field in one with the suitable equation of
state for a cosmological constant, mimicking the solid state physics and without introducing a quintessence field. 
The model in \cite{1} is also capable of explaining why only a positive cosmological constant generally emerges
and not one with $\overline{\Lambda}<0$.
Finally, the smallness of the cosmological constant can be explained in terms of a decoherence scale $L_D$ representing
an absolute minimum for the generalized
Misner-Sharp mass dressed by Planckian fluctuations and the crossover to classicality can be obtained with a large class of initial conditions. For all these results, I think that my proposal is physically viable and can represent a 
possible, simple and  physically reasonable solution to the cosmological constant problem. It is thus expected that the 
semiclassical treatment 
and the physical mechanisms outlined in this paper and in \cite{1} can be present in a future physically sound quantum gravity theory,
where particles and quantum fluctuations are rigorously depicted in terms of quantum fields.

\end{document}